\documentclass[10pt,fleqn]{article}

\usepackage{amssymb}

\newcommand{\name}[1]{\begin{flushleft}
                       \LARGE \bf #1
                       \end{flushleft}\vspace{-3mm}}

\newcommand{\Author}[1]{\begin{flushleft}
                       \it #1 \end{flushleft}}

\newcommand{\Adress}[1]{\begin{flushleft}
                       \it #1 \end{flushleft}}

\newcommand{\AbsEng}[1]{
    \begin{flushright}
    \begin{minipage}{120mm}
     \small   #1
    \end{minipage}
    \end{flushright}
}

\newcommand{\be}{\begin{equation}}
\newcommand{\ee}{\end{equation}}
\newcommand{\ba}{\hspace*{-5pt}\begin{array}}
\newcommand{\ea}{\end{array}}
\newcommand{\p}{\partial}
\newcommand{\ds}{\displaystyle}

\newcommand{\pbf}[1]{\mbox{\mathversion{bold}$#1$}}

\begin{document}

\name{Equations of motion in odd-dimensional spaces and
${\pbf T}$-, ${\pbf C}$-invariance}

\medskip

\noindent{published as Preprint of Institute of Theor. Phys., N 69-17, Kyiv, 1969, 13~p.}

\Author{Wilhelm I. FUSHCHYCH}

\Adress{Institute of Mathematics of the National Academy of
Sciences of Ukraine,\\ 3 Tereshchenkivska  Street, 01601 Kyiv-4,
UKRAINE}

\noindent {\tt URL:
http://www.imath.kiev.ua/\~{}appmath/wif.html\\ E-mail:
symmetry@imath.kiev.ua}

\AbsEng{The properties of the equation of Dirac type in three-dimensional and five-dimensional
Minkowski space-time with respect to time reflection (in sense of Pauli and Wigner) as
well as to the operation of charge conjugation are investigated. $P$-, $T$-, $C$-invariance
of Dirac equation for the cases of four components (in three-dimentional space) and
eight components (in five-dimensional space) is established. Within the framework of
the Poincar\'e group a relativistic equation is suggested wich describes the movement
of a particle with non-fixed (indefinite) mass in external electromagnetic field.}

\medskip

\centerline{\bf Introduction}

F. KIein and latter de Broglie pointed out the usefulness of spaces with more than four
dimensions for the construction of the physical theories. This idea was intensively developed
in 1930--1940 years by many authors who tried to unify the gravitation and electromagnetic
theories. Nowadays it is widely developed in connection with the extension of the
Poincar\'e group  $({\mathcal P}(1,3))$ as well as with idea of combining ${\mathcal P}(1,3)$
with group of internal symmetries (a review of this works can be found in~[1]).

In the works [2] the mass operator was proposed to be defined as one like mo\-men\-tum
or angular momentum operator, i.e. we proposed to define the mass operator to be not
a Casimir operator but the generator of a group wich has the Poincar\'e group as its subgroup.
For such a group in papers~[3,~4] the inhomogeneous de Sitter, group is chosen --- a group
of rotations and translations in 5-dimensional flat Minkowski space-time with the square-mass
operator beings related to the generator $P_4$ (of group ${\mathcal P}(1,4)$)
in such a way
\[
M^2 =\varkappa^2 +P_4^2.
\]

In the present work the $P$-, $T$-, $C$-invariance properties of the simplest equations
invariant under the group ${\mathcal P}(1,4)$  are investigated.

\medskip

\centerline{\bf \S~1. Dirac equation within $\pbf{{\mathcal P}(1,4)}$ scheme}
\centerline{\bf and $\pbf{P}$-, $\pbf{T}$-, $\pbf{C}$-transformations}

The simplest equations invariant under ${\mathcal P}(1,4)$ group are Dirac equations wich
in the Hamilton form can be writen down as following:
\renewcommand{\theequation}{1.\arabic{equation}}
\setcounter{equation}{0}
\be
H^+\Psi^+(t,\vec x)=i \frac{\p \Psi^+ (t,\vec x)}{\p t},
\ee
\be
H^-\Psi^-(t,\vec x)=i \frac{\p \Psi^- (t,\vec x)}{\p t},
\ee
\be
\ba{l}
\ds H^\pm \equiv \alpha_k p_k\pm \beta \varkappa, \qquad p_k =-i \frac{\p}{\p x_k},
\qquad k=1,2,3,4,
\vspace{2mm}\\
\alpha_k =\gamma_0 \gamma_k, \qquad \beta=\gamma_0, \qquad
\vec x\equiv (x_1, x_2, x_3. x_4),
\ea
\ee
where $\gamma_\mu$  are five four-dimensional Dirac matrices $(\mu=0,1,2,3,4)$.

The invariance of equation (1.1) (or (1.2)) under space-inversion $x_k\to -x_k$ is obvious
sinse in $(1+4)$-dimensional Minkowski space-time this inversion is reduced to a rotation.

Let us clear up now the question of the invariance of the equation~(1.1) (or~(1.2)) under the
time reflection $(t\to -t)$ and charge conjugation. To this aim we write down the generators
of the group ${\mathcal P}(1,4)$ defined of the solutions of the equations~(1.1) and~(1.2)
explicitely
\be
\ba{l}
P_0 =H^+ , \qquad P_k=p_k,
\vspace{2mm}\\
\ds J_{kl} =x_k p_l -x_l p_k +\frac i2 \alpha_l \alpha_k,
\vspace{2mm}\\
\ds J_{0k} =x_0 p_k -\frac 12 (x_k P_0+P_0 x_k),
\vspace{2mm}\\
\ds [x_k,p_l]=i\delta_{kl}, \qquad [x_k, x_l]=[p_k, p_l] =0.
\ea
\ee

According to Pauli the time-reflection operator  $T^p$ satisfies the conditions
\be
T^p \Psi (t,\vec x) =\tau^p \Psi (-t,\vec x), \qquad \left(T^p\right)^2=1,
\ee
\be
[T^p, P_0]_+=0, \qquad [T^p, P_k]_+=0, \qquad [T^p, J_{kl}]_-=0,
[T^p, J_{0k}]_+=0,
\ee
where  $\tau^p$ is a $(4\times 4)$-matrix.

According to Wigner the time-reflection operator $T^w$ must satisfy the following conditions
\be
T^w \Psi (t,\vec x) =\tau^w \Psi^* (-t,\vec x), \qquad \left(T^w\right)^2=1,
\ee
\be
[T^w, P_0]=0, \qquad [T^w, P_k]_+=0, \qquad [T^w, J_{kl}]_+=0,
[T^w, J_{0k}]=0,
\ee
where  $\tau^w$ is a   $(4\times 4)$ matrix.

\renewcommand{\thefootnote}{\arabic{footnote}}
\setcounter{footnote}{0}

Finaly the charge-conjugation operator must satisfy the
conditions\footnote{In general the squares of operators $T^p$, $T^w$ and $C$
are equal to unity to within a multiplicative factor of unit modulus.}
\be
C\Psi(t,\vec x)=\tau^c \Psi^* (t,\vec x), \qquad C^2=1,
\ee
\be
[C, P_0]_+=[C,P_k]_+=0, \qquad [C, J_{\mu\nu}]_+=0,
\ee
where  $\tau^c$   is a $(4\times 4)$ matrix.

Matrices $\tau^p$, $\tau^w$ and $\tau^c$ can be representated in folowing form
\be
\tau^p =a_\mu^p \alpha_\mu +a^p_{\mu\nu} \alpha_\mu \alpha_\nu, \qquad \mu <\nu,
\ee
\be
\tau^w =a_\mu^w \alpha_\mu +a^w_{\mu\nu} \alpha_\mu \alpha_\nu, \qquad \mu <\nu,
\ee
\be
\tau^c =a_\mu^c \alpha_\mu +a^c_{\mu\nu} \alpha_\mu \alpha_\nu, \qquad \mu <\nu,
\ee
where $a_\mu$, $a_{\mu\nu}$ are the arbitrary numbers $(\mu=0,1,2,3,4)$.

Using (1.11) and (1.13) one can immediately verify that the relations~(1.6) and (1.10)
are satisfied only for the zero-matrices $\tau^p$ and $\tau^c$. Relation~(1.7) is satisfied if
$\tau^w=\alpha_1 \cdot \alpha_3$.

Thus, the equation (1.1) or (1.2) is $T^p$-, $C$-noninvariant but $P$, $T^w$-invariant.
This means that the four-component Dirac equations in five-dimensional scheme are not
$PTC$-invariant as it was pointed out in~[4,~5].

This result is a consequence of the fact that in contrary to the usual Dirac equation~(1.1)
(or~(1.2)) do not describe a particle and antiparticle. In fact the generators of the group
${\mathcal P}(1,4)$ given in the form~(1.4) defined on the manifold of all solutions of
equations~(1.1) and~(1.2) realize the representations
\be
D^+(1/2,0)\oplus D^-(0,1/2),
\ee
\be
D^+(0,1/2)\oplus D^- (1/2,0)
\ee
respectively. As it is commonly known, the usual Dirac equation describes
a particle and antiparticle and on the manifold of all its solutions the representation
$D^+(1/2)\oplus D^-(1/2)$ is realized of group ${\mathcal P}(1,3)$.

Starting from the equation (1.1) (or (1.2)) and using Bargman--Wigners method~[6] one can
describe some class of equations invariant under the ${\mathcal P}(1,4)$
group and the time reflection in sense of Wigner, however they are noninvariant under
$T^p$ and $C$ operations.

Hence we see that (1.1) (or (1.2)) as well as the class of the Bargman-Wigner type equations
(derived from~(1.1) o~ (1.2)) are $T^w$-invariant, but $T^p$-, $C$-noninvariant.

It may seen in this connection that any theory wich is built up in five-dimensional
Minkowski space-time is always $PTC$-noninvariant~[5]. Though actualy it is not so.
In fact, let us consider equation
\be
H\Psi(t,\vec x) =i \frac{\p \Psi(t,\vec x)}{\p t} , \qquad
\Psi(t) \equiv \Psi(t,\vec x)=\left( \begin{array}{c}
\Psi^+(t,\vec x)
\vspace{1mm}\\
\Psi^-(t,\vec x)\end{array}\right),
\ee
where
\be
\ba{l}
H\equiv \widetilde \alpha_k p_k +\widetilde \beta \varkappa, \qquad k=1,2,3,4,
\vspace{1mm}\\
\widetilde \alpha_k =
\left( \begin{array}{cc}
\alpha_k & 0\\
0 & \alpha_k \end{array} \right),
\qquad
\widetilde \beta =
\left( \begin{array}{cc}
\beta & 0\\
0 & -\beta \end{array} \right).
\ea
\ee
On the manifold of solutions of this equations operators $T^w$, $T^p$ and $C$ are defined as:
\be
T^p\Psi (t) = \widetilde \tau^p \Psi(-t), \qquad
T^w\Psi (t) = \widetilde \tau^w \Psi^*(-t), \qquad
C\Psi (t) = \widetilde \tau^c \Psi^*(t),
\ee
\be
\widetilde \tau^p =\left(\begin{array}{cc}
0 & \beta\\
\beta & 0
\end{array} \right), \quad
\widetilde \tau^w =\left(\begin{array}{ccc}
\alpha_1 & \alpha_3  & 0\\
 0 & \alpha_1 & \alpha_3 \end{array} \right), \quad
\widetilde \tau^c =\left(\begin{array}{ccc}
0 & \alpha_2 & \alpha_4 \\
\alpha_2 & \alpha_4  &0 \end{array} \right).
\ee

One can immediatly verify that the relations (1.16), (1.8) and~(1.10) actually satisfy for the
equation~(1.16). In means that the equation~(1.16) is $T^w$-, $T^p$-, $C$-  and
$PTC$-invariant. That is also clear from the fact that equation~(1.16) realizes representation
\be
D^+(1/2,0) \oplus D^-(1/2,0)\oplus D^+(0,1/2) \oplus D^-(0,1/2).
\ee

Starting from (1.16) and generalizing the Bargman--Wigner method on ${\mathcal P}(1,4)$
group one can describe all the equations of Bargman--Wigner type wich are
$PTC$-invariant~[7].

Thus in case of five dimensions one has to choose for the basic equation on eight-component
equation~(1.16) but not a four-component equation~(1.1) or~(1.2).

If cane puts in (1.1) $\varkappa=0$, then such four-component equation is
$T^p$-, $C$-invariant and in this case:
\be
\tau^p=\gamma_0, \qquad \tau^c=\alpha_2\alpha_4, \qquad \tau^w=\alpha_1\alpha_3.
\ee
Equation (1.1)
\renewcommand{\theequation}{1.\arabic{equation}${}'$}
\setcounter{equation}{0}
\be
\alpha_k p_k \Psi^\pm (t,\vec x)=i \frac{\p \Psi^\pm (t,\vec x)}{\p t}
\ee
describes a particle whose spin is $1/2$ but the mass is non-fixed since
\renewcommand{\theequation}{1.\arabic{equation}}
\setcounter{equation}{21}
\be
M^2 \widetilde \Psi (t,\vec p\,) =p_4^2 \widetilde \Psi(t,\vec p\,)=m^2 \widetilde \Psi(t,\vec p\,),
\qquad -\infty< p_4 <\infty, \quad 0\leq m^2 \leq \infty.
\ee
Here $\widetilde \Psi(t,\vec p\,)$ is the Fourier-image of function $\Psi(t,\vec x)$.

From what was performed above it reveales that in ${\mathcal P}(1,4)$
scheme it is possible to describe a particle with non-fixed mass (i.e. the particles of
resonance type) the spin of wich fixed.

\bigskip

\centerline{\bf \S~2. Equation for a particle with non-fixed mass on
$\pbf{{\mathcal P}(1,3)}$ group}

In this section we show how one can write down the relativistic equation of mation for a
particle with the non-fixed mass within the framework of Poincar\'e group.
Usually elementary particle either stable or unstable whose spin is $s$,
is associated with a Hilbert space $R^s(m)$ in wich on irreducible representation of the
Poincar\'e group ${\mathcal P}(1,3)$ is realized. Such a correspondence is unjustified one
since we cannot attribute the definite mass to the unstable particle. Following~[2,~8] let us
attribute to an unstable particle (resonance) a Hilbert space $R^s$ with is the direct integral
of spaces $R^s(m)$, i.e.
\renewcommand{\theequation}{2.\arabic{equation}}
\setcounter{equation}{0}
\be
R^s =\int \oplus R^s (m) g^s(m^2) dm^2,
\ee
where function  $g^s(m^2)$ is not equal to zero only within the interval
$[m_1^2, m_2^2]$ wich characterizes the spread (indefinite) of mass of a particle.

According to (2.1) each vector from $R^s$ can be representated as
\be
\ba{l}
\ds \Psi^s(t,\vec x) =\int \oplus \Psi ^s (t,\vec x, m) g^s (m^2) dm^2,
\vspace{2mm}\\
\ds \Psi^2(t,\vec x,m)\in R^s(m), \qquad \vec x\equiv (x_1,x_2,x_3),
\ea
\ee
\be
P^2_\mu \Psi^s (t,\vec x,m) =m^2 \Psi^s (t,\vec x,m), \qquad \mu=0,1,2,3,
\ee
\be
P^2_\mu\Psi^s(t,\vec x)=\int\oplus m^2 \Psi^s (t,\vec x,m) g^s(m^2) dm^2.
\ee

The generators of the Poincar\'e group on vectors (2.2) are defined in such a way
\be
P_\mu\Psi^s(t,\vec x)=\int\oplus P_\mu \Psi^s(t,\vec x,m) g^s(m^2) dm^2,
\ee
\be
J_{\mu\nu} \Psi^s(t,\vec x)=\int\oplus J_{\mu\nu} \Psi^s (t,\vec x,m) g^s(m^2) dm^2.
\ee
The Dirac equation for the function  $\Psi^{s=1/2}(t,\vec x)$ is:
\be
\left( i\gamma_0 p_0+i \gamma_k p_k -\sqrt{p_\mu^2}\right) \Psi^{s=1/2}(t,\vec x)=0.
\ee
One can easily see now that (2.7) can be reduced to the usual Dirac equation if one
formaly replaces the function $g^{s=1/2}(m^2)$ in~(2.2) by $\delta(m^2-m_0^2)$.
The generators of ${\mathcal P}(1,3)$ group defined by~(2.5) and~(2.6) on the manifold of
solutions of eq.~(2.7) are given by~(1.4), where
\be
P_0\equiv H\equiv \alpha_k p_k +\beta\sqrt{p_\mu^2}.
\ee

We can write down the equation of motion for a particle with indefinite mass, wich interacts
with the external electromagnetic field in form
\be
\left( i\gamma_0 \pi_0 +i\gamma_k \pi_k -\sqrt{\pi_\mu^2}\right) \Psi^{s=1/2}(t,\vec x)=0,
\ee
where $\pi_\mu \equiv p_\mu -e A_\mu$. It is clear that equation~(2.9) essentialy differs
from the usual Dirac equation wich discribes the motion of a particle with fixed mass in
the electromagnetic field. A detaled analysis if equation~(2.9) will be performed in a
forthcoming work.

Lurcat [8] pointed out, that interpretation of function $\Psi^s(t,\vec x)$ as a wave function
of particle is not correct.

More appropriate is to characterize the unstable system by the density
matrix (operator). In the Schr\"odinger picture the equation of motion for the density matrix
looks like
\[
i\frac{\p \rho}{\p t}=[H,\rho],
\]
where $H$ is defined by (2.8).

Equation (2.7) as well as the usual Dirac equation, is $P$-, $T$-, $C$-invariant.

\bigskip

\centerline{\bf \S~3. Equation for the flat particle and $\pbf{T}$-, $\pbf{C}$-invariance}

To clear up how can extend the obtained above (sec. 1) results upon any arbitrary group
${\mathcal P}(1,2n+1)$ let us consider in this section equations of motion wich are
invariant under ${\mathcal P}(1,2)$ group (the group of rotations and translations
in three-dimensional Minkowski space).

The simplest equations invariant under ${\mathcal P}(1,2)$ are:
\renewcommand{\theequation}{3.\arabic{equation}}
\setcounter{equation}{0}
\be
H^+\Psi^+ (t,x_1,x_2) =i\frac{\Psi^+(t,x_1,x_2)}{\p t},
\ee
\be
H^-\Psi^- (t,x_1,x_2) =i\frac{\Psi^-(t,x_1,x_2)}{\p t},
\ee
\be
\ba{l}
H^\pm =\alpha_kp_k \pm \beta \varkappa, \qquad k=1,2,
\vspace{1mm}\\
\ds \alpha_1=\sigma_1, \qquad \alpha_2 =\sigma_2, \qquad \beta=\sigma_3,
\qquad p_k =-i\frac{\p}{\p x_k},
\ea
\ee
Here $\Psi(t,x_1,x_2)$ is a two-component spinor, and $\sigma_1$, $\sigma_2$,
$\sigma_3$ are Pauli manrices.

Taking into account that in this case
\be
\tau^p=a^p\cdot 1 +\vec a^{\,p}\vec \sigma, \qquad
\tau^w=a^w\cdot 1 +\vec a^{\,w} \sigma, \qquad
\tau^c=a^c\cdot 1 +\vec a^{\,c} \sigma
\ee
and arguing in a way similar to that of sec. 1, we reveal that equation~(3.1) or~(3.2) is
$T^p$-, $T^w$- and $PTC$-noninvariant but $P$- and  $C$-invariant.

Equation
\be
\ba{l}
\ds H\Psi (t,x_1,x_2) =i\frac{\p \psi (t,x_1,x_2)}{\p t},
\vspace{2mm}\\
\ds \Psi(t) \equiv \Psi(t,\vec x) =\left( \begin{array}{c}
\Psi^+ (t,\vec x)
\vspace{1mm}\\
\Psi^-(t,\vec x) \end{array}\right), \qquad H=\widetilde \alpha_k p_k +\widetilde \beta \varkappa,
\qquad k=1,2,
\ea
\ee
\be
\widetilde \alpha_k =\left( \begin{array}{cc}
\alpha_k & 0\\
0 & \alpha_k \end{array}\right), \qquad
\widetilde \beta =\left( \begin{array}{cc}
\sigma_3 & 0\\
0 & \sigma_3 \end{array}\right)
\ee
is $T^p$-, $T^w$- and $C$-invariant as well as equation~(1.16) is, i.e. it is
$PTC$-invariant, and for matrices and $\tau^p$, $\tau^w$ we have
\be
\widetilde \tau^p =\left( \begin{array}{cc}
0 &\sigma_3 \\
 \sigma_3  & 0 \end{array}\right), \qquad
\widetilde \tau^w =\left( \begin{array}{cc}
0 &\sigma_2 \\
 \sigma_2  & 0 \end{array}\right), \qquad
\widetilde \tau^c =\left( \begin{array}{cc}
\sigma_1 & 0 \\
0&  \sigma_1 \end{array}\right).
\ee

Thus equations of (1.1), (3.1) type and a whole class of equations of Bargman--Wigner
type wich are derived from the equations of~(1.1) type are invariant under the limited groups:

 ${\mathcal P}(1,2)$ are $T^p$-, $T^w$-, $T^wC$-noninvariant and $C$-invariant;

 ${\mathcal P}(1,4)$ are $T^p$-, $C$-, $T^wC$-noninvariant and $T^w$-, $T^pC$-invariant;

  ${\mathcal P}(1,6)$ are $T^p$-, $T^w$-, $T^pC$-, $T^wC$-noninvariant and $C$-invariant;

${\mathcal P}(1,8)$ are $T^p$-, $T^w$-, $T^wC$-noninvariant and $T^w$-, $T^pC$-invariant.

$\vdots$

To prove the assertiona given above in the case of arbitrary ${\mathcal P}(1,2n+1)$
group one has to carry out the very similar procedure to that we employed for ${\mathcal P}(1,4)$
group and to use the fact that Dirac matrices $\gamma^{(2n+1)}$ of group
${\mathcal P}(1,2n)$ are related with those $\gamma^{(2n-1)}$ of group
${\mathcal P}(1,2n-2)$ by
\[
\ba{l}
\ds \left( \gamma_\mu^{(2n+1)}, \gamma_{2n}^{(2n+1)}, \gamma_{2n+1}^{(2n+1)}\right)=
\left( \gamma_\mu^{(2n-1)} \otimes \sigma_2, 1\otimes \sigma_3, 1\otimes \sigma_1\right),
\vspace{1mm}\\
\mu=0,1,\ldots,2n-1.
\ea
\]

Puting in (3.1) and (3.2) $\varkappa=0$ one sees that equation~(3.1) concides with~(3.2)
and such equation is $C$-, $T$-invariant, and $\tau^p=\sigma_3$, $\tau^w=\sigma_2$.

\medskip

\begin{enumerate}

\footnotesize

\item Hegerfeldt G.C., Henning J., {\it  Fortschr. Phys.}, 1968, {\bf 16}, 9.

\item Fushchych W.I., {\it Ukr. Phys. J.}, 1968, {\bf 13}, 256; 1967, {\bf 12},
741.\ \  {\tt quant-ph/0206056}

\item Fushchych W.I., Krivsky I.Yu., {\it Nucl. Phys. B}, 1968, {\bf 7},
79.\ \  {\tt quant-ph/0206057}

\item Fushchych W.I., Kryvski I.Yu., Preprint ITF-68-72, Kyiv, 1968.

\item Rosen S.P., {\it J. Math. Phys.}, 1968, {\bf 9}, 1593.

\item Bargman Y., Wigner E., {\it Proc. Nat. Acad. Sci.}, 1948, {\bf 34}, 211.

\item Fushchych W.I., Sokur L.P., Preprint ITF-69-33,  Kyiv, 1969.

\item Lurcat F., {\it Phys. Rev.}, 1968, {\bf 173}, 1461.

\item Brauer B., Weyl H., {\it  Am. J. Math.}, 1935, {\bf 57}, 447.

\end{enumerate}
\end{document}